\documentclass[a4paper,twoside]{article}

\usepackage{epsfig}
\usepackage{url}
\usepackage{subcaption}
\usepackage{calc}
\usepackage{cite}
\usepackage{amssymb}
\usepackage{amstext}
\usepackage{amsmath}
\usepackage{amsthm}
\usepackage{enumitem}
\usepackage{float}
\usepackage{multicol}
\usepackage{pslatex}
\usepackage{natbib}
\usepackage{algorithm2e}
\usepackage[bottom]{footmisc}
\usepackage{SCITEPRESS}  
\usepackage{graphicx}

\begin{document}

\title{Adapting Under Fire: Multi-Agent Reinforcement Learning for Adversarial Drift in Network Security}
\author{
\authorname{
Emilia Rivas\sup{1}, 
Sabrina Saika\sup{1}, 
Ahtesham Bakht\sup{2}, 
Aritran Piplai\sup{1}, 
Nathaniel D. Bastian\sup{3}, 
Ankit Shah\sup{4}
}
\affiliation{\sup{1}The University of Texas at El Paso, USA}
\affiliation{\sup{2}University of South Florida, USA}
\affiliation{\sup{3}United States Military Academy, USA}
\affiliation{\sup{4}Indiana University, USA}
\email{\{erivas6, ssaika\}@miners.utep.edu, apiplai@utep.edu, ahthesham@usf.edu, nathaniel.bastian@westpoint.edu, ankit@iu.edu}
}
\vspace{-0.5em}
\keywords{Network Intrusion Detection System, Adversarial Attack, Drift Adaptation, Reinforcement Learning}
\vspace{-0.5em}
\abstract{Evolving attacks are a critical challenge for the long-term success of Network Intrusion Detection Systems (NIDS). The rise of these changing patterns has exposed the limitations of traditional network security methods. While signature-based methods are used to detect different types of attacks, they often fail to detect unknown attacks. Moreover, the system requires frequent updates with new signatures as the attackers are constantly changing their tactics. In this paper, we design an environment where two agents improve their policies over time. The adversarial agent, referred to as the red agent, perturbs packets to evade the intrusion detection mechanism, whereas the blue agent learns new defensive policies using drift adaptation techniques to counter the attacks. Both agents adapt iteratively: the red agent responds to the evolving NIDS, while the blue agent adjusts to emerging attack patterns. By studying the model's learned policy, we offer concrete insights into drift adaptation techniques with high utility. Experiments show that the blue agent boosts model accuracy by ~30\% with just 2–3 adaptation steps using only 25–30 samples each.}

\onecolumn \maketitle \normalsize \setcounter{footnote}{0} \vfill

\section{\uppercase{Introduction}}
\label{sec:introduction}
 
Network Intrusion Detection Systems (NIDS) play a critical role in monitoring and identifying malicious traffic. However, they are increasingly challenged by adversarial attacks, where inputs are intentionally crafted to evade detection. Separately, data drift —shifts in network packet patterns over time— can degrade detection performance if models are not regularly updated. 

In a real-world network security system, a packet passing through the network may be adversarial and capable of executing an attack. NIDS monitor traffic for suspicious behavior and alert the security team when something abnormal is detected. The security team then investigates and attributes certain packets as sources of the attack. However, this human-in-the-loop process introduces delays, often allowing the attack to succeed before any defensive action is taken. To reduce this delay, automatic adaptation mechanisms are essential—an area where machine learning (ML) models play a critical role. Yet, as defenders adopt ML to improve response time and detection accuracy, adversaries are also evolving their strategies, using ML techniques to evade detection, further complicating the defense landscape. 
 
Consider a mid-sized financial organization where, one afternoon, a wave of obfuscated packets bypasses the ML-enabled NIDS, triggering no immediate alerts. Hours later, analysts in the Security Operations Center (SOC) investigate a spike in failed transactions and retroactively label several earlier packets as malicious. The team now faces a key decision: \textit{Should they retrain the NIDS using a specific drift adaptation technique or invest additional time and resources into identifying the most informative, uncertain samples?} Given limited time and computational resources, understanding which adaptation strategy—or combination thereof—is most effective in restoring detection performance becomes a critical research and operational concern.
 
In prior works such as DeepPackGen \citep{hore2025deep}, deep reinforcement learning (DRL) has demonstrated strong performance in generating perturbations in samples that successfully evade a NIDS. Similarly, Chale et al. proposed a constrained optimization method for generating adversarial examples in network intrusion detection systems, which ensures that perturbed packets remain functional under protocol constraints \citep{chale2024constrained}. However, both approaches assume that their defender model is static, which in most cases is not valid in real-world scenarios. In practice, organizations periodically update their defense mechanism to adapt to the changing attacks. Additionally, the available data is typically not accurately labeled, making the adaptation process more challenging.

In this paper, our goal is to model a multi-agent drift adaptation game where one player- the red agent- will execute valid network perturbations, whereas the other player- the blue agent- will perform drift adaptation by adjusting to the distribution shift introduced by perturbed samples. The key contributions of this paper are three-fold:
\begin{itemize}[leftmargin=*, noitemsep]
    \item We create a novel RL environment that unifies packet-level attacks and drift adaptation into a single co-evolving game between attacker and defender agents. We will make our code publicly available.
    \item To our knowledge, this is the first agent-based drift adaptation framework for NIDS that integrates efficient adaptation techniques, such as active and continual learning.
    \item We model intrusion detection as a long-term multi-agent process that captures strategic evolution. 
\end{itemize}

\vspace{-20pt}
\section{\uppercase{Related Works}}
\vspace{-5pt}
\noindent\textbf{Adversarial Attacks in Network Security:} 
Adversarial robustness has made significant advances in recent years, particularly in the areas of malware analysis and intrusion detection. Despite the progress in defense mechanisms, existing methods still cannot perform well under evolving attacks \citep{bai2021recent}.

\noindent\textbf{Intrusion Detection System: }
Traditional intrusion detection systems often fail to cope with changing network attacks. 
Recent studies have proposed adaptive techniques such as ensemble learning, online retraining, and drift detection to mitigate these issues
\citep{jaw2021feature,shyaa2024evolving}. However, most of these methods treat adaptation as a periodic, reactive process, without addressing real-time adversarial perturbations.

\noindent\textbf{Drift Adaptation Techniques:} The drift method has been widely studied in adversarial settings using methods for detection. Surveys by Lu et al. and Gama et al. provide comprehensive foundations for drift adaptation techniques in non-stationary environments \citep{lu2014concept, gama2014survey}.
Active learning and semi-supervised learning have been widely adopted for cost reduction of manual labeling and to help classifiers adapt when labeled data is limited or uncertain\citep{settles2009active}. Techniques such as uncertainty sampling, clustering-based selection, and core-set methods have shown strong performance across various domains~\citep{ouali2020overview}. 
Continual learning and contrastive learning offer promising solutions to mitigate catastrophic forgetting and improve model resilience in the face of evolving threats. These techniques have shown effectiveness in tasks such as Android malware detection and intrusion classification~\citep{yue_contrastive_2022,wu_contrastive_2022}.

\noindent\textbf{Multi-Agent Reinforcement Learning:} Multi-agent reinforcement learning (MARL) has also been explored in network security for dynamic defense mechanisms and optimal resource allocation.
 Previous works demonstrate that in many security-based scenarios MARL performed better than the traditional security techniques. Zhou et al.  provide a detailed overview of the application of MARL across different domains, including network security. Along with the overview, the author mentions that MARL solves the limitations of the traditional NIDS through collaborative learning. However, its relevance in adaptive adversarial situations is limited because the majority of existing works either assume cooperative agent settings or ignore the concept drift \citep{zhou2023multi}.

\noindent\textbf{MARL with Drift Adaptation:} Previously, Kuppa et al.\citep{kuppa2022learn} proposed a detection mechanism based on drift techniques where their method is static and focused on detecting drifted or adversarial samples and discovering new classes. Shayesteh et al. also proposed a static drift model that relies on traditional methods like transfer and ensemble learning for drift detection, closely related to our work \citep{9718523}. 

\noindent\textbf{Cyber Security Simulation Environment:} Several cybersecurity simulation environments have been proposed, each addressing different aspects of network defense. For instance - Hammar et al. \citep{hammar2020finding} proposed self play and reinforcement learning to generate reasonable security strategies in a simplified Markov game. Similarly, the CyberBattleSim platform emphasizes system-level access and privilege escalation \citep{msft:cyberbattlesim}. Building on such environments, Piplai et al. introduced a knowledge-guided two-player reinforcement learning framework that incorporates cybersecurity knowledge graphs to train both attacker and defender agents, demonstrating improved learning efficiency and reduced network disruption compared to traditional setups\citep{10068930}. 

\par Despite notable progress in adversarial robustness, adaptive NIDS, and MARL applications, current methods either treat adaptation statically, assume benign agent behaviors, or neglect dynamic adversarial drift. Our work addresses these gaps by introducing a dynamic, dual-agent reinforcement learning system capable of continuous drift adaptation under adversarial perturbations. To our knowledge, this is the first integration of MARL and real-time drift adaptation techniques specifically targeted at evolving network threats within a dedicated simulation environment.

\vspace{-20pt}
\section{\uppercase{Methodology}}
\vspace{-10pt}
\noindent\textbf{Framework Overview:} Our framework consists of three primary components —a  machine learning-based classifier, a red agent training module, and a blue agent training module— and was applied in two separate experiments using the CICIDS-2017 and ACI IoT datasets, respectively.

\noindent\textbf{Datasets and Preprocessing:} We conduct our experiment using the CICIDS-2017 dataset \citep{sharafaldin2018toward}. It contains both benign and attack samples in the form of raw pcap files. The data is captured from the period 3 July 2017 (9 am Monday) to the 7 July 2017 (5 pm Friday). The identified attack types in this data collection are Brute Force FTP, Brute Force SSH, DoS, Heartbleed, Web Attack, Infiltration, Botnet, and DDoS.

We also used the ACI IoT Network Traffic Dataset 2023 \citep{qacj-3x32-23} for our experiment. The ACI IoT environment replicates the operational conditions of a standard home Internet of Things (IoT) network, incorporating a heterogeneous mix of both wired and wireless devices. The dataset utilized in this research employs a multi-modal data representation, encompassing network traffic patterns, inter-device communication, and device-specific attributes. Furthermore, this dataset emphasizes the behavioral analysis of IoT devices by examining intricate network behaviors under both normal operational states and adversarial conditions. The dataset consists of the following labelled classes: Benign DNS Flood, Dictionary Attack, Slowloris, SYN Flood, Port Scan, Vulnerability Scan, OS Scan, UDP Flood, and ICMP Flood. The experiment for data collection was conducted in 2023 from October 30th to November 3rd.

Preprocessing begins with accessing the corresponding dataset, which adheres to the TCP/IP model for network communication. The raw network traffic data is stored in the libpcap (pcap) format, which is widely recognized as the de facto standard for network packet capture. We follow appropriate packet labeling guidelines. Each TCP packet contains up to 1,594 bytes of information; however, not all of these bytes are utilized, as certain protocol-related and header-specific information can introduce bias into the model. Consequently, the Ethernet (ETH) header, source and destination IP addresses, as well as source and destination port numbers, were removed from each packet.

A packet may contain a maximum payload size of 1,460 bytes, but not all packets reach this limit. To maintain a standardized data structure, packets with fewer than 1,460 bytes of payload data were zero-padded. Following preprocessing, the extracted feature length is 1,525. Finally, the raw packet information, originally represented in hexadecimal format, was converted to decimal values ranging from 0 to 255 and subsequently normalized to [0, 1] to improve computational efficiency and ensure that learning operates in a continuous input space.


\noindent\textbf{Red Agent and Adversarial Sample Generation:} Within the proposed framework, the objective of the red agent is to generate adversarial perturbations on malicious network packets from the datasets, enabling them to evade detection by machine learning-based classifiers while preserving their original malicious functionality. This process involves introducing a perturbation ‘$\delta$’ to a malicious packet such that the modified packet, $x_p$, is misclassified as benign by the defender’s classifier. Formally, $x_p$ is defined as:

$x_p = P(x_{original}, \delta)$

\noindent where $P$ represents the perturbation function applied to the original packet. The adversarial packets in this study are generated following the approach outlined in \citep{hore2025deep}.

The process of generating adversarial samples for network packets can be formulated as a sequential decision-making problem. In this framework, a reinforcement learning agent learns to apply perturbations iteratively using deep reinforcement learning (DRL). At each iteration, the agent modifies the packet in a manner that maximizes a predefined reward function, thereby increasing the likelihood of evading detection by the defender model. This problem is modeled as a Markov Decision Process (MDP), with the following key components:

\begin{itemize}[leftmargin=*,noitemsep]
    \item \textbf{State:} The state represents the normalized byte values of the network packet obtained from the preprocessed dataset.
    \item \textbf{Action:} The action space consists of valid perturbations, $\delta \in \Delta$ that the agent can apply at each time step $t$.
    \item \textbf{Reward:} The reward function quantifies the effectiveness of taking action $a_t$ at $t$. It is designed to guide the agent toward learning an optimal policy that maximizes the likelihood of deceiving the defender model while maintaining packet function.
\end{itemize}

\begin{figure}
    \centering
    \includegraphics[width=1.1\linewidth]{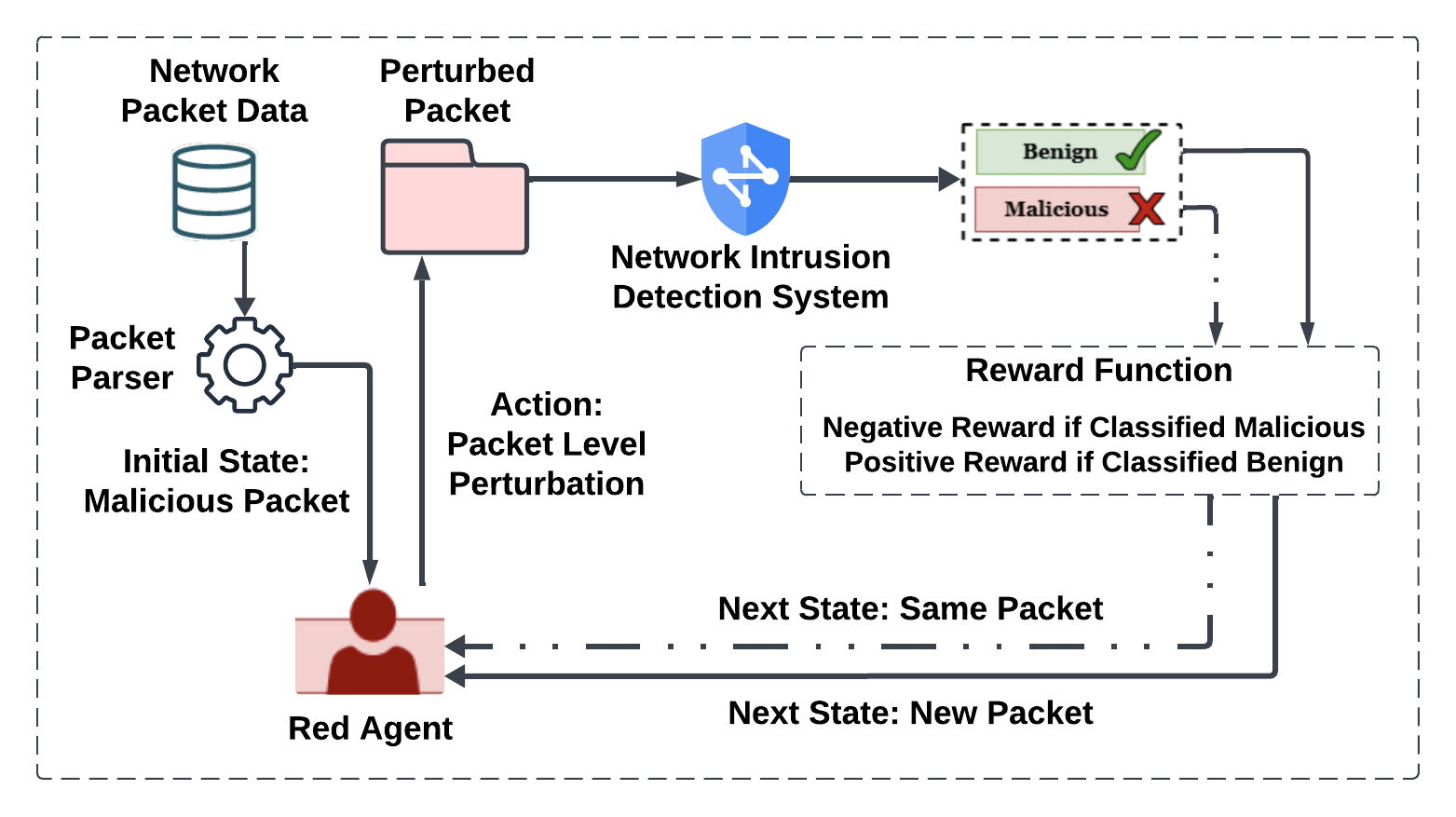}
    \caption{Red Agent Training of the DRL agent}
    \label{Red Team}
\end{figure}

\noindent\textbf{DRL Training: } The training environment for the deep reinforcement learning (DRL) agent consists of labeled network packets and pre-trained classifiers from the initial stage. The environment's state is defined as a randomly selected malicious network packet. Each action taken by the agent results in a state transition, generating a perturbed sample. A novel reward function is designed to guide the agent toward learning an optimal policy for evading the classifier. The process continues iteratively until the adversarial sample is successfully misclassified or a predefined maximum number of steps is reached. Figure \ref{Red Team} shows the red team training of the DRL agent. During this phase the agent learns an optimal policy for generating the adversarial samples.

To train the red agent, we employ DRL with Deep Q-Network (DQN), Proximal Policy Optimization (PPO) and Advantage Actor Critic (A2C) . DQN and PPO  is well-suited for problems with discrete action spaces, whereas A2C can be used in both continuous and discrete spaces. 
The perturbation techniques were applied based on the general principles of preserving packet functionality while introducing adversarial noise, as discussed in prior works.
\citep{nasr2021defeating,sadeghzadeh2021adversarial,guo2021black, apruzzese2020deep,huang2020adversarial}. Specifically, the modifications included:

\begin{itemize}
\item Modifying the fragmentation bytes from \\``Do Not Fragment" to ``Fragment."
\item Modifying the fragmentation bytes from ``Do Not Fragment" to ``More Fragments."
\item Increasing or decreasing the Time-To-Live (TTL) byte value.
\item Adjusting the window size bytes by increasing or decreasing their values.
\item Modifying the maximum segment size (MSS) value by adding, increasing, or decreasing it.
\item Adjusting the window scale value by adding, increasing, or decreasing it.
\item Incorporating additional segment information into the packet.
\end {itemize}
These perturbations were carefully designed to maintain packet integrity while enhancing the adversarial capability to evade detection.

\vspace{1em}

\noindent\textbf{Blue Agent: } 
The goal of the blue agent is to protect the network from attack by selecting optimal drift adaptation techniques. Here, the blue agent mainly uses four drift adaptation techniques to adapt and update its policy over time. Using these techniques, the agent learns how and when to update itself. When new data arrives, the agent chooses these techniques, where each technique represents an action. The action allows the agent to update its classifier based on the technique it has chosen. In each batch, multiple drift adaptation techniques can be used. The state, action, and reward function of the blue agent is defined as follows- 
\newline
\newline
\textbf{State:}  The state in blue agent is represented as the difference of the mean value of the features between what the model has seen and what the model has yet to see. \par 

 Let \( \mu_{\text{new}} \) be the mean vector of the current batch, and \( \mu_{\text{seen}} \) be the mean vector of all previously seen data. The feature-difference \( f_t \) at time \( t \)  can be defined as follows:

\vspace*{-.4cm}
\begin{equation}
f_t = \mu_{\text{new}} - \mu_{\text{seen}} = \frac{1}{n_{\text{new}}} \sum_{i=1}^{n_{\text{new}}} x_i^{\text{new}} - \frac{1}{n_{\text{seen}}} \sum_{j=1}^{n_{\text{seen}}} x_j^{\text{seen}}
\end{equation}

\vspace*{-.3cm}
where:
\begin{itemize}
    \item \( x_i^{\text{new}} \) is the \( i \)-th sample in the current batch of size \( n_{\text{new}} \),
    \item \( x_j^{\text{seen}} \) is the \( j \)-th sample from all previously observed data (size \( n_{\text{seen}} \)).
\end{itemize}

 In addition to the feature difference, the state vector contains five metrics: classification accuracy, false positive rate (FPR), false negative rate (FNR), KL divergence to measure distributional shift, and Wasserstein distance to quantify the statistical difference between current and previously seen data. Thus, the final state vector is defined as:
\begin{equation}
s_t = [\text{Acc}, \text{FPR}, \text{FNR}, D_{\text{KL}}, D_{\text{W}}, f_t]
\label{eq:state_vector}
\end{equation}


\vspace{1em}

\noindent\textbf{Action:} The state changes based on the four drift adaptation techniques (Active learning, Online Learning, Continual Learning, Pseudo Labeling) which are actions for the blue agent. 
From the four discrete actions, the blue agent can select any of the actions to update its state and can adapt its policy .These strategies help the agent to be robust to the changes when the adversaries changes. Preferably more than one action is needed to choose in a single batch for adaptation. The description of the four types of actions are as follows: \par

\noindent\textbf{Action 0: Online Learning:} In this strategy \citep{shalev2012online}, the RL algorithm interacts with the environment in real time to gradually update the model by using the most recent batch. The benefit of interacting in real time is that there is no need to retrain from scratch for updates. Given a set of samples from the environment, we randomly select a subset \(\mathcal{Q}\) of size \(B\) which is defined by the query budget. 


\noindent\textbf{Action 1: Active  Learning:} This  focuses on the most informative data rather than all of the samples\citep{settles2009active}. It selectively queries the label to update the model. For this learning model, we need to calculate the uncertainty using the entropy of the predicted class probabilities. To perform active learning under a query budget, we use uncertainty sampling by using the model's predicted class probabilities. For each incoming sample \(x_i\), we compute the prediction confidence as the maximum class probability, and select those within a predefined uncertainty range \([p_{\text{low}}, p_{\text{high}}]\). If the number of uncertain samples exceeds the query budget \(B\), we select the top-\(B\) most uncertain samples (i.e., those closest to 0.5 confidence). 

\noindent\textbf{Action 2: Continual Learning:} Here the agent learns knowledge over time without additional training on the full dataset. The key idea of this approach is that it gives more weight to recent samples than the older ones.The formula is defined as follows:
\vspace{-.1cm}
\begin{equation}
\mathcal{R} = \mathcal{R}_{\text{rep}} \cup \mathcal{R}_{\text{disc}}
\label{eq:replay_split}
\end{equation}

\noindent
\text{where}
\vspace{-.25cm}
\[
\mathcal{R}_{\text{rep}} = \{x_i \in \mathcal{B} \mid \mathcal{H}(x_i) < \tau_{\text{low}}\}
\]
\[
\mathcal{R}_{\text{disc}} = \{x_i \in \mathcal{B} \mid \mathcal{H}(x_i) > \tau_{\text{high}}\}
\]

\noindent
Here, \(\mathcal{B}\) is the input sample set, \(\mathcal{H}(x_i)\) is the entropy of sample \(x_i\), and \(\tau_{\text{low}}\), \(\tau_{\text{high}}\) are the entropy thresholds used to select representative and discriminative samples, respectively. We use a combination of these samples, $\mathcal{R}$, to adapt the model. We use a setup that performs well on security setting \citep{rahman2025madar}.

\noindent\textbf{Action 3: Pseudo Labeling } This is a semi-supervised learning technique where a model selects unlabeled samples with prediction confidence above a threshold and treats the predicted labels as if they were true \citep{arazo2020pseudo}. These pseudo-labeled samples are then used to refine the model in the absence of sufficient labeled data. Let \(\theta\) denote the model parameters, \(f_\theta(x_i)\) the model prediction, and \(\hat{y}_i = \arg\max f_\theta(x_i)\) the assigned pseudo-label. The model is updated to reduce the loss over the set \(\mathcal{P}\) of selected pseudo-labeled samples:
\[
\text{Update}(\theta; \mathcal{P}) \quad \text{to minimize} \quad \sum_{(x_i, \hat{y}_i) \in \mathcal{P}} \mathcal{L}(f_\theta(x_i), \hat{y}_i)
\]
This update may or may not involve gradient-based methods, depending on the learning framework.

\vspace{5pt}
\noindent\textbf{Reward:} Multiple actions can be taken on a single batch, with each action transitioning the environment to a new state and yielding a reward. The reward structure for the blue agent is designed to balance accuracy improvement with sample efficiency.

Let \( \text{Acc}_{\text{test}}^t \) denote the model's test accuracy at time \( t \), and \( \text{Acc}_{\text{test}}^{t-1} \) be the accuracy at the previous time step. Let \( r = \frac{|\mathcal{D}_{\text{used}}|}{|\mathcal{D}_{\text{train}}|} \) represent the ratio of training samples used in the current step, relative to the total available training data. We also define \( T \) as a fixed test accuracy threshold. The reward \( R_t \) is then defined as:

\vspace{-.4cm}
\begin{small}
\begin{equation}
R_t = 
\begin{cases}
10, & \text{if } \text{Acc}_{\text{test}}^t > T \\
-10 \cdot r + 50 \cdot (\text{Acc}_{\text{test}}^t - \text{Acc}_{\text{test}}^{t-1}), & \text{otherwise}
\end{cases}
\end{equation}
\end{small}

The constants used in this reward formulation are chosen to enforce a trade-off between performance and efficiency. The flat reward of \( 10 \) when the accuracy exceeds the threshold \( T \) serves as a consistent incentive to reach and maintain high performance without unnecessary retraining. The penalty term \( -10 \cdot r \) discourages excessive data usage by assigning a cost proportional to the fraction of the training set used. This helps promote data-efficient behavior, particularly important under concept drift where retraining can be costly. The improvement term \( 50 \cdot (\text{Acc}_{\text{test}}^t - \text{Acc}_{\text{test}}^{t-1}) \) rewards actual gains in test performance, encouraging the agent to select actions that lead to meaningful learning. The multiplier 50 ensures that even small improvements (e.g., 1–2\%) yield noticeable rewards, helping to counterbalance the penalty

\vspace{-20pt}
\section{\uppercase{Experimental Results}}
\vspace{-5pt}
In our experimental setup, we model a two-player interaction between a red agent (attacker) and a blue agent (defender) to simulate a dynamic adversarial environment. The red agent aims to compromise the intrusion detection system (IDS) by perturbing data samples, while the blue agent responds by adapting its defense strategy. Both agents learn and update their respective policies in alternating rounds, allowing for a continual adaptation process. This iterative framework enables the defender to stay resilient against evolving and previously unseen attack strategies, reflecting real-world conditions.\par
\begin{figure}[h]
    \centering
    \includegraphics[width=1.05\linewidth]{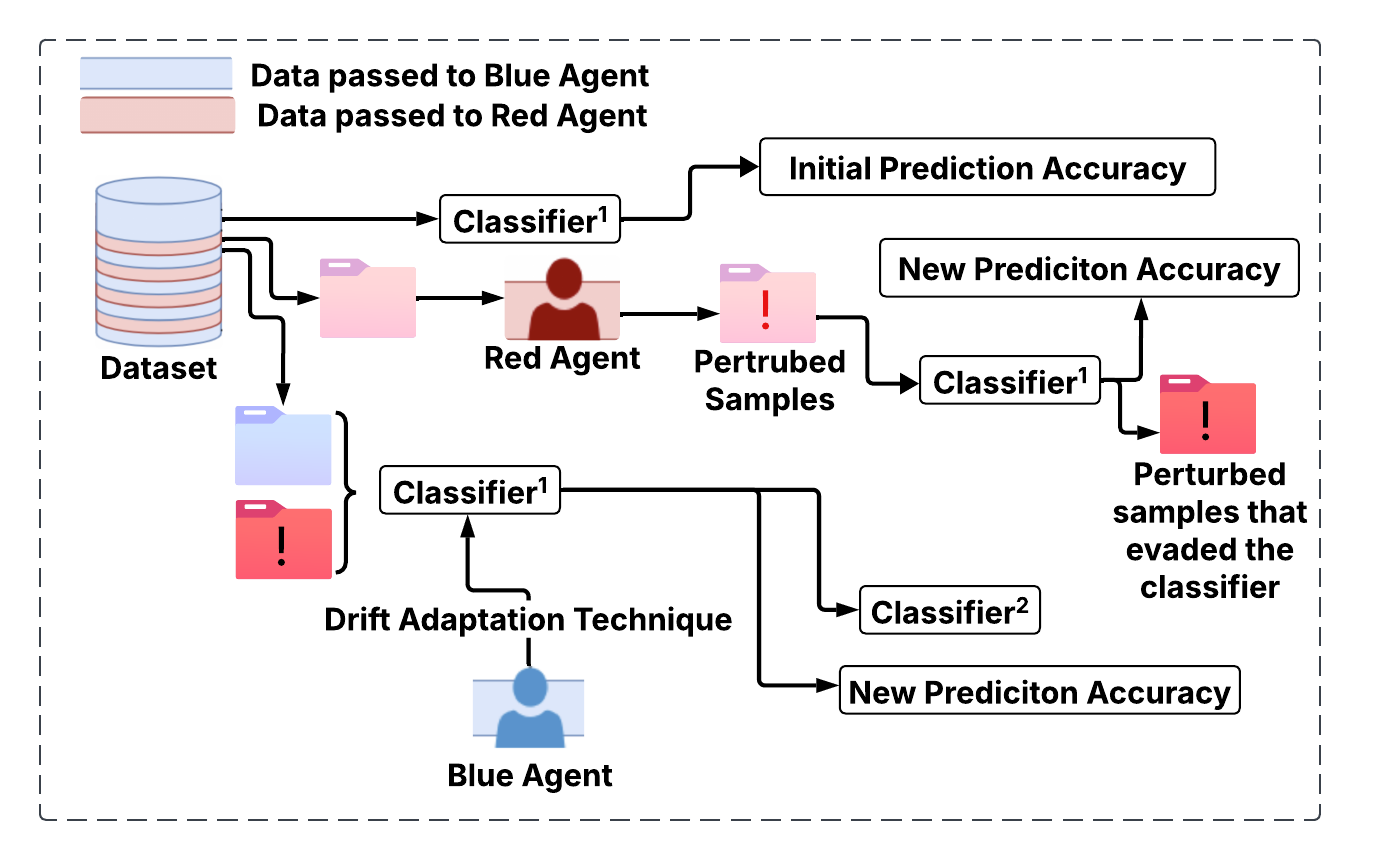} 
    \caption{Overview of the adaptive defense framework using CICIDs 2017 and ACI IoT datasets}
    \label{fig:weighting}
\end{figure}

We divide the dataset into 120 sequential batches, where each batch is used alternately by the red and blue agents to update their respective policies. We describe this setup in Figure \ref{fig:weighting}. During the red agent’s turn, it treats the NIDS (classifier) as static and attempts to perturb the current batch of data to evade detection. The success of the red agent is measured by its ability to generate perturbations that bypass the existing classifier. In the subsequent turn, the blue agent receives both the red agent’s perturbed data and the new batch of data, using them to retrain and update the NIDS model. This updated classifier is then made available to the red agent for its next turn, allowing it to adapt its perturbation strategy based on the most recent defense model. Through this alternating interaction, the red agent continuously learns to attack a moving target, while the blue agent incrementally hardens the classifier using drift adaptation techniques. This two-player adaptation technique helps both the agents to adapt to the changing policies.

\begin{figure}[h!]
    \centering
    \hspace*{-0.3cm}
    \includegraphics[width=1.05\linewidth]{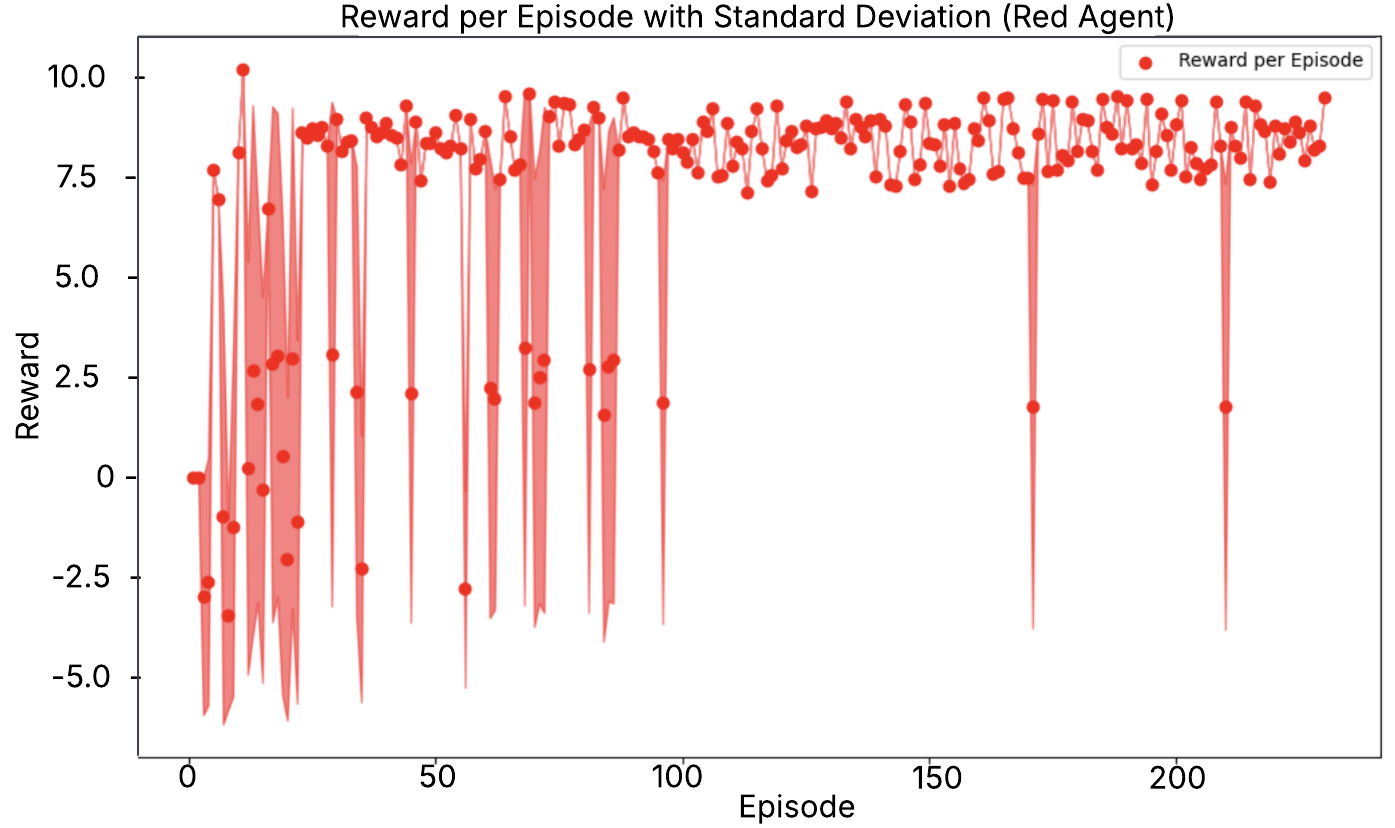}
    \caption{Episodic rewards and standard deviation during Red Agent training (SAC) , reflecting its effectiveness in perturbing models to evade classification.}
    \label{fig:episodic-rewards-red}
\end{figure}

\begin{figure}
    \centering
    \includegraphics[width=1\linewidth]{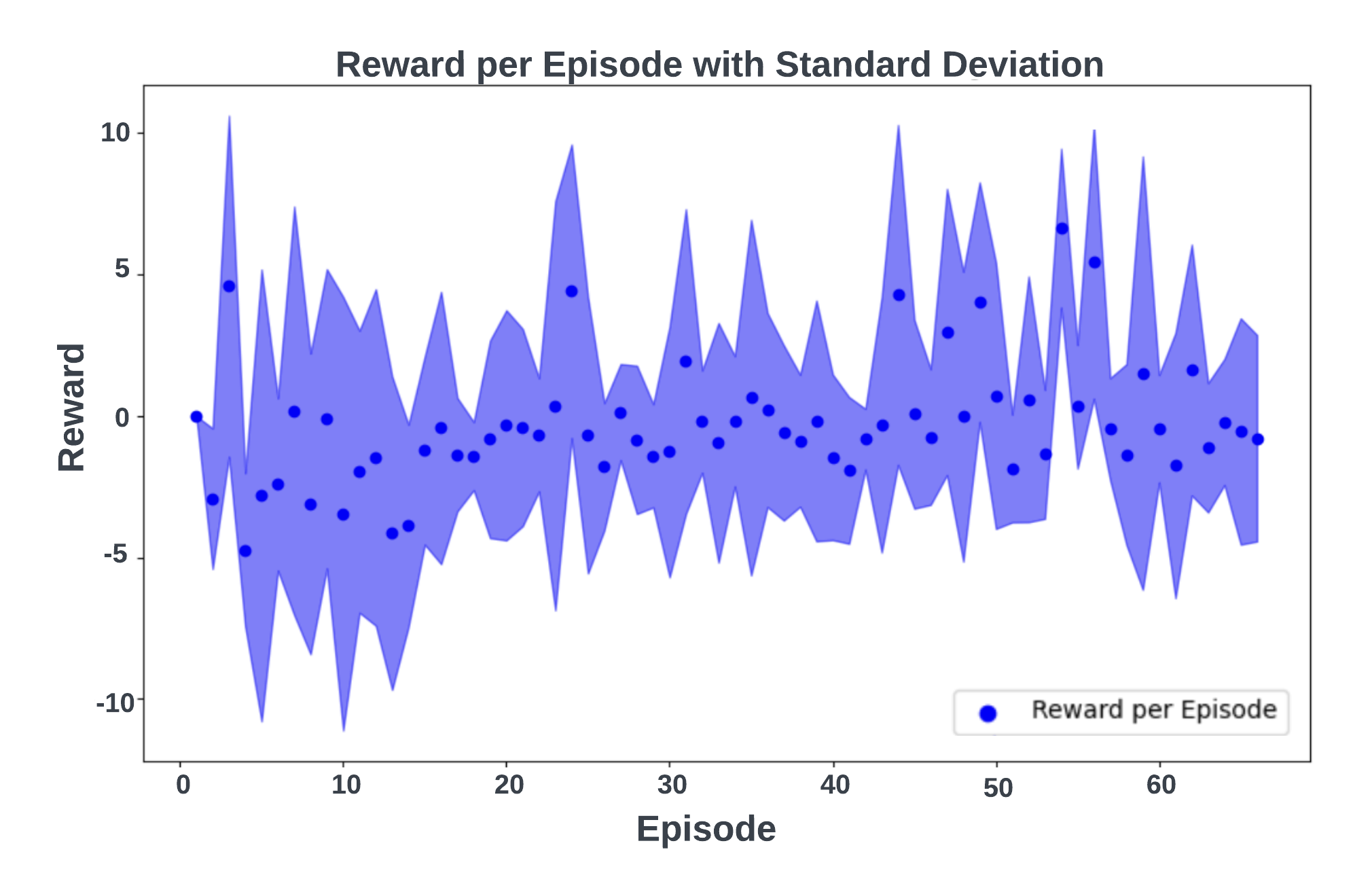}
    \caption{Episodic rewards and standard deviation during Blue Agent training(DQN). The dots represent the average episodic reward.}
    \label{fig:episodic-rewards-blue}
\end{figure}

\noindent\textbf{Results:} We perform several evaluations to offer insights into the training for the red and the blue agents. In Figures \ref{fig:episodic-rewards-red} and \ref{fig:episodic-rewards-blue}, we observe the mean episodic rewards for the red and the blue agent and the standard deviations. 

We observe the mean and standard deviation of the red agent’s rewards and find that the model trains relatively quickly. The low standard deviation and high mean observed after 100 episodes suggest that the red agent is able to successfully evade the classifier using only a small number of actions.

As shown in Figure \ref{fig:episodic-rewards-blue}, the episodic rewards for the blue agent exhibit a slower but more stable increase over time. The standard deviation is notably higher compared to the red agent, indicating that not all actions yield consistent outcomes. Due to the exploratory nature of the reinforcement learning agent, suboptimal actions are more frequent during the early stages of training but tend to diminish as the agent converges toward a more effective policy.

\begin{figure}[h!]
    \centering
    \includegraphics[width=1.1\linewidth]{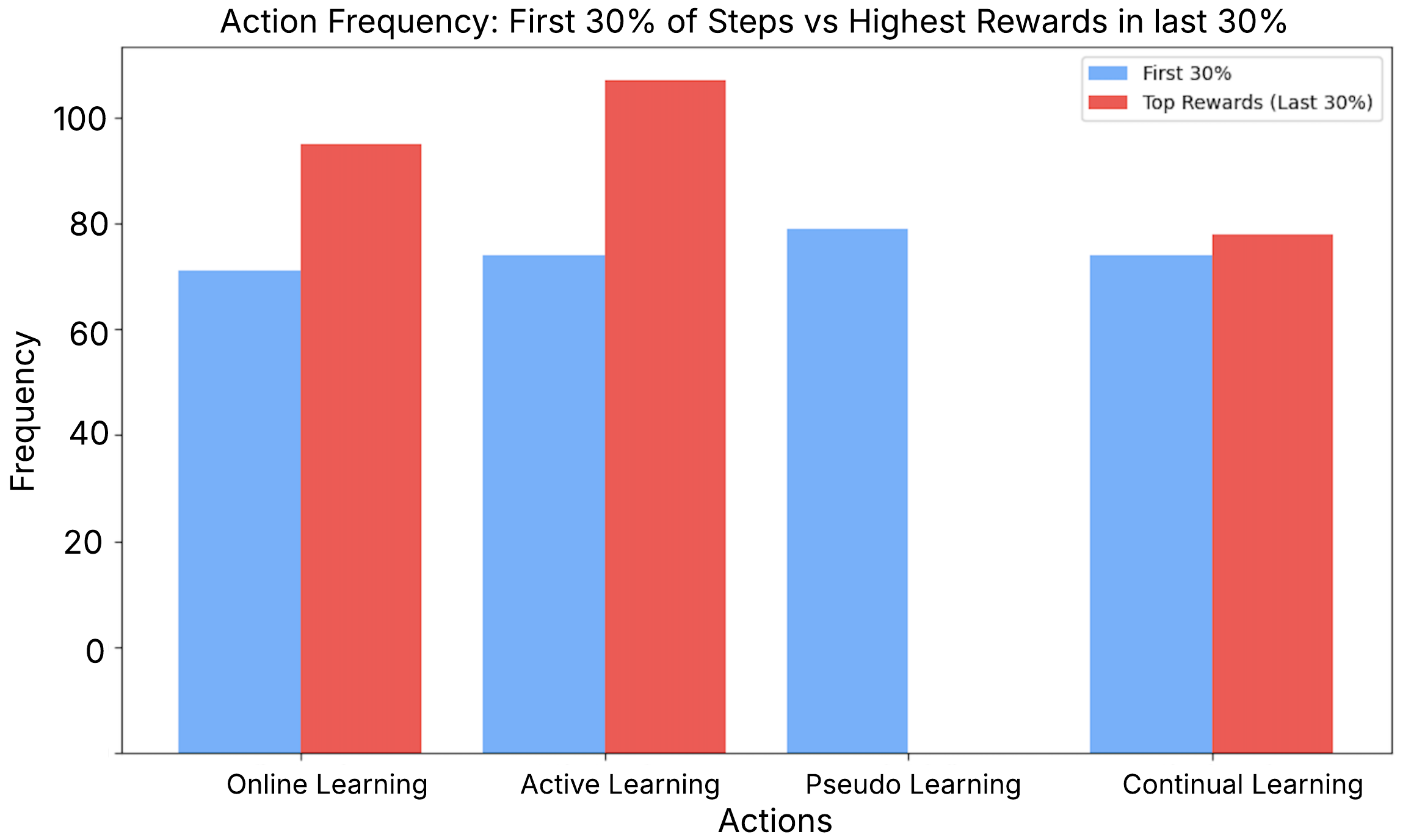}
    \caption{Action frequency distribution during Blue Agent training.}
    \label{fig:enter-label}
\end{figure}

The computational cost of the blue agent’s actions is significantly higher than that of the red agent, as they involve updating XGBoost models, which are computationally intensive and difficult to parallelize efficiently. Despite the higher cost, we deliberately chose XGBoost due to its widespread use in the industry and its strong, reliable performance across a variety of classification tasks, particularly in security-related applications.

\begin{table}[ht]
\centering
\begin{tabular}{|c|p{2.5cm}|}
\hline
\textbf{Action} & \textbf{Average Reward} \\
\hline
Active Learning & \hspace{.7cm}0.2157 \\
\hline
Online Learning & \hspace{.7cm}0.1526 \\
\hline
Pseudo Labeling & \hspace{.6cm}-4.6639 \\
\hline
Continual Learning & \hspace{.6cm}-1.4697 \\
\hline
\end{tabular}
\caption{Average rewards for each action during Blue Agent training.}
\label{tab:average_rewards}
\end{table}

Tables \ref{tab:red-performance-cic}, \ref{tab:red-performance-aci} show the drop in classifier performance after the red agent's training, both before and after the blue agent's response. The change in accuracy reflects how the classifier is impacted by the artificial drift introduced through the red agent’s perturbations. Both the CIC-IDS 2017 and ACI-IoT datasets yield high accuracy on malware and benign classes. While the processed CIC-IDS 2017 dataset is balanced, the ACI-IoT dataset exhibits class imbalance. We include ACI-IoT in our experiments not only to assess generalizability but also because CIC-IDS 2017—though widely used—has become a standard benchmark, whereas ACI-IoT is a more recently collected, timely dataset. 

\begin{table}[h!]
\centering
\resizebox{0.5\textwidth}{!}{%
\begin{tabular}{|l|l|l|l|}
\hline
    & All state features & \begin{tabular}[c]{@{}l@{}}State features\\ without models'\\ characteristics\end{tabular} & \begin{tabular}[c]{@{}l@{}}State features\\ without data\\ characteristics\end{tabular} \\ \hline
DQN & \hspace{.7cm}89.34 &\hspace{.7cm}87.18 & \hspace{.7cm}72.46 \\ \hline
PPO & \hspace{.7cm}92.05&\hspace{.7cm}86.96 &\hspace{.7cm}84.81 \\ \hline
A2C & \hspace{.7cm}87.34 & \hspace{.7cm}85.89 & \hspace{.7cm}75.44\\ \hline
\end{tabular}%
}
\caption{Ablation study to understand the impact of state representation when learning the appropriate sequence of drift adaptation. The scores reflect the average final accuracy on adapting to a new dataset. Initial accuracies ranged between 62-65\%. }
\label{tab:blue-statistics}
\end{table}

\vspace{-.4cm}
\begin{table}[h!]
\centering
\resizebox{0.5\textwidth}{!}{%
\begin{tabular}{|l|l|l|}
\hline
     & \begin{tabular}[c]{@{}l@{}}Change in accuracy\\ (before blue action)\end{tabular} & \begin{tabular}[c]{@{}l@{}}Change in accuracy\\ (after blue action)\end{tabular} \\ \hline
SAC  & \hspace{.3cm}97.08 - 64.46\%  &\hspace{.3cm}91.89 - 78.48\% \\ \hline
DDPG & \hspace{.3cm}97.50 - 64.33\% &\hspace{.3cm}93.72 - 77.57\%  \\ \hline
PPO  & \hspace{.3cm}96.25 - 69.44\% & \hspace{.3cm}96.12 - 75.88 \%  \\ \hline
\end{tabular}%
}
\caption{Drop in Model Accuracy for CIC-IDS caused by red agent }
\label{tab:red-performance-cic}
\end{table}

\vspace{-.4cm}
\begin{table}[h!]
\centering
\resizebox{0.5\textwidth}{!}{%
\begin{tabular}{|l|l|l|}
\hline
     & \begin{tabular}[c]{@{}l@{}}Change in accuracy\\ (before blue action)\end{tabular} & \begin{tabular}[c]{@{}l@{}}Change in accuracy\\ (after blue action)\end{tabular} \\ \hline
SAC  & \hspace{.3cm}97.62 - 68.46 \%  & \hspace{.3cm}86.07 - 68.01\% \\ \hline
DDPG & \hspace{.3cm}94.59 - 69.65\%   & \hspace{.3cm}79.62 - 66.21 \% \\ \hline
PPO  & \hspace{.3cm}93.46 - 69.50\%   & \hspace{.3cm}85.92 - 68.35\%  \\ \hline
\end{tabular}%
}
\caption{Drop in \textit{Balanced} Accuracy for ACI-IoT dataset caused by red agent }
\label{tab:red-performance-aci}
\end{table}

In Table \ref{tab:blue-statistics}, we conduct an ablation study to identify the most effective state representation for our blue drift adaptation environment, hypothesizing that an autonomous agent can select an appropriate adaptation strategy based on the classifier’s status and the data distribution. The baseline state includes both model-specific and data-specific characteristics (Section 3), and our results show that data characteristics generally have a greater influence on the agent’s decision. To further analyze the agent’s behavior, we examine how its policy evolves during training by comparing the distribution of selected actions over time (Figure \ref{fig:enter-label}). We find that reliance on pseudo-labeling declines, suggesting reduced reliability under drift, while online and active learning become dominant strategies. Continual learning also contributes, though it requires identifying samples from the known distribution to populate a replay buffer—an added cost not captured in our reward function, which is based solely on test performance improvement and sample efficiency.

\vspace{-20pt}
\section{\uppercase{Conclusion}}
\label{sec:conclusion}
\vspace{-5pt}
Our setup reflects real-world dynamics, where adversaries continually evolve, and defenders must regularly update or refresh their models. In practice, harmful packets may initially go undetected and are only labeled as malicious or benign after further analysis by a SOC, often post-incident. This delayed feedback highlights the importance of selecting effective drift adaptation techniques. Our findings suggest that randomized online learning can be surprisingly effective, even with as few as 20 samples. Active learning, which focuses on querying uncertain samples, also shows promise; however, it may falter when the model is overly confident, resulting in too few samples being selected for meaningful updates. Therefore, a practical recommendation for organizations, especially under time constraints, is to use a simple approach with small number of random samples, as it offers a strong balance between speed and effectiveness. 

We present a novel reinforcement learning environment that simulates the co-evolution of network intrusion and defense strategies through two interacting agents: an adaptive attacker and a learning defender. By framing intrusion detection as a multi-agent process, we demonstrate that drift adaptation techniques, especially those requiring minimal labeled data, can significantly enhance detection performance, achieving up to a 30\% accuracy improvement. 
We analyze attacker behavior following defender adaptation and explore optimal state space representations for the adapting defender. 
In future, we plan to investigate scenarios with greater temporal drift and conduct long-range analyses incorporating a broader variety of evolving attack types to more accurately reflect the dynamic threat landscape faced by modern NIDS.

\vspace{10pt}
\begin{small}
\bibliography{example}
\bibliographystyle{apalike}

\end{small}

\end{document}